\documentclass{article}
\usepackage{amssymb}
\usepackage{amsfonts}
\usepackage{graphicx}
\usepackage{amsmath}
\usepackage{geometry}
\usepackage{scalefnt}
\usepackage{ulem}
\usepackage{hyperref}
\usepackage{slashed}
\usepackage{multicol}

\input{tcilatex}

\begin{document}

\title{Deformation Quantization: From Quantum Mechanics to Quantum Field
Theory}
\author{Philip Tillman \\
%EndAName
Department of Physics and Astronomy, University of Pittsburgh, Pittsburgh,
PA, USA}
\date{\today }
\maketitle

\begin{abstract}
The aim of this paper is to give a basic overview of Deformation
Quantization (DQ) to physicists. A summary is given here of some of the key
developments over the past thirty years in the context of physics, from
quantum mechanics to quantum field theory. Also, we discuss some of the
conceptual advantages of DQ and how DQ may be related to algebraic quantum
field theory. Additionally, our previous results are summarized which
includes the\textit{\ }construction of the Fedosov star-product on dS/AdS.
One of the goals of these results was to verify that DQ gave the same
results as previous analyses of these spaces. Another was to verify that the
formal series used in the conventional treatment converged by obtaining 
\textit{exact} and \textit{nonperturbative} results for these spaces.
\end{abstract}

%TCIMACRO{\TeXButton{\begin{multicols}{2}}{\begin{multicols}{2}}}%
%BeginExpansion
\begin{multicols}{2}%
%EndExpansion
%TCIMACRO{\TeXButton{\vspace{-15pt}}{\vspace{-15pt}}}%
%BeginExpansion
\vspace{-15pt}%
%EndExpansion

\section{Introduction}

%TCIMACRO{\TeXButton{\vspace{-15pt}}{\vspace{-15pt}}}%
%BeginExpansion
\vspace{-15pt}%
%EndExpansion
There are three standard approaches to quantum theories: the operator
formalism, the path integral formalism, and deformation quantization (DQ).
The aim of this proceedings is to inform the reader of the state of the DQ
in terms of issues relevant to physicists: from quantum mechanics to quantum
field theory.

The main problem in DQ is the issue of convergence of all perturbative
series which remain unknown. To address this issue, I will summarize some of
the results of $\left[ \text{TiSp1}\right] $ which includes the computation
of the Fedosov star-product (the fundamental object used in DQ) exactly for
the dS and AdS space-times. Another goal of the results were to reproduce
previous results for the Klein-Gordon (KG) equation on dS and AdS.$\left[ 
\text{Fro1-4}\right] $

The question is: Why bother with DQ at all? The reason we do is that DQ
provides some distinct advantages over canonical quantization and the path
integral methods. One example is that it is not only coordinate invariant
but also independent of the choice of dynamics (e.g. Lagrangian). The
associativity of the star-product plays a fundamental role in understanding
how DQ deviates from canonical quantization as can be seen in $\left[ \text{%
Til,GoRe}\right] $.

In this paper we will discuss other advantages. For instance, the
observables in DQ are functions on phase-space just as they are classically.
Therefore, the conceptual break with classical mechanics is less severe than
with, for example, operator methods which map observables to fundamentally
different objects. Most of the tools used for functions, whether they are
geometric or algebraic, extend much more naturally into DQ than in other
quantization methods.

Furthermore, it is argued in $\left[ \text{D\"{u}Fr}\right] $ that
perturbative algebraic quantum field theory (AQFT) can be understood in
terms of DQ. It is argued that this is the likely scenario because of the
many similarities between the two approaches. Conventional treatments of DQ
rely heavily on perturbative expansions in the formal parameter $\hbar $
which makes it easy to obtain perturbative results for various physical
quantities. However, besides the perturbative nature of DQ, AQFT focuses on
the algebra of observables just as DQ does. Other similarities to AQFT are
noticed like the way in which the topology of ordinary phase-space functions
induces observable topology in DQ to form the "net of observables" in AQFT.$%
\left[ \text{Haag,Buc}\right] $

%TCIMACRO{\TeXButton{\vspace{-15pt}}{\vspace{-15pt}}}%
%BeginExpansion
\vspace{-15pt}%
%EndExpansion

\section{An Introduction to 
%TCIMACRO{\TeXButton{\\}{\\ }}%
%BeginExpansion
\\ %
%EndExpansion
Deformation Quantization}

%TCIMACRO{\TeXButton{\vspace{-15pt}}{\vspace{-15pt}}}%
%BeginExpansion
\vspace{-15pt}%
%EndExpansion
In 1927 Herman Weyl wrote his quantization rule $\mathcal{W}$ that maps
every phase-space function $f$ on a flat space-time to a unique observable
in the space of linear operators acting on the appropriate Hilbert space.
Shortly afterwards, Eugene Wigner wrote the inverse map $\mathcal{W}^{-1}$.$%
\left[ \text{Weyl,Wig,DiSt}\right] $ Groenewold in 1946 $\left[ \text{Gro}%
\right] $ (and later Moyal in 1949 $\left[ \text{Moy}\right] $) investigated
the formula $\mathcal{W}^{-1}\left( \mathcal{W}\left( f\right) \mathcal{W}%
\left( g\right) \right) $ and found a remarkable result:%
\begin{eqnarray}
&&\mathcal{W}^{-1}\left( \mathcal{W}\left( f\right) \mathcal{W}\left(
g\right) \right)  \label{Moyal0} \\
&=&f\exp \left[ \frac{i\hbar }{2}\left( \frac{\overleftarrow{\partial }}{%
\partial x^{\mu }}\frac{\overrightarrow{\partial }}{\partial p_{\mu }}-\frac{%
\overleftarrow{\partial }}{\partial p_{\mu }}\frac{\overrightarrow{\partial }%
}{\partial x^{\mu }}\right) \right] g  \notag
\end{eqnarray}%
where the operator inside the exponential is the Poisson Bracket and the
arrows over the derivatives explain the direction in which they act.

Moreover, Groenewold (and again later Moyal) realized that this operator is
an associative, noncommutative product of the two phase-space functions $f$
and $g$ defined by $f\ast g:=\mathcal{W}^{-1}\left( \mathcal{W}\left(
f\right) \mathcal{W}\left( g\right) \right) $ which has the familiar
commutators:%
\begin{equation*}
\left[ x^{\mu },p_{\nu }\right] _{\ast }=i\hbar \delta _{\nu }^{\mu }\text{
\ \ },\text{ \ \ }\left[ x^{\mu },x^{\nu }\right] _{\ast }=0=\left[ p_{\mu
},p_{\nu }\right] _{\ast }
\end{equation*}

In a coordinate independent formulation we have:%
\begin{equation}
f\ast g=f\exp \left[ \overleftrightarrow{P}\right] g  \label{Moyal}
\end{equation}%
\begin{equation*}
\overleftrightarrow{P}:=\overleftarrow{\partial }_{A}\frac{i\hbar }{2}\omega
_{AB}\overrightarrow{\partial }_{B}
\end{equation*}%
where $\overleftrightarrow{P}$ is the Poisson bracket and $\partial _{A}$ is
a (flat) torsion-free phase-space connection ($\partial \otimes \omega =0$).

In summary, they obtained another equivalent formulation of the quantum
theory on phase-space.$\left[ \text{HWW,HiHe}\right] $

Despite all of this, it wasn't until 1978 when Bayen F., Flato M., Fr\o %
nsdal C., Lichnerowicz A., Sternheimer D. proposed an alternative
formulation of quantum theory on the phase-space of an arbitrary curved
space-time that we now know as DQ. The new formulation (and a quite radical
one) can aptly be summarized by a quote from their paper:

"\textit{We suggest that quantization be understood as a deformation of the
structure of the algebra of classical observables, rather than as a radical
change in the nature of the observables}"-Bayen et al. (1978)

The view here is that quantum mechanics is a theory on a classical
phase-space by the replacement of the pointwise product by a new associative
but noncommutative star-product. This star-product is a pseudodifferential
operator i.e. an operator of the form:

%TCIMACRO{\TeXButton{\end{multicols}}{\end{multicols}}}%
%BeginExpansion
\end{multicols}%
%EndExpansion
%TCIMACRO{\TeXButton{\vspace{-15pt}}{\vspace{-15pt}}}%
%BeginExpansion
\vspace{-15pt}%
%EndExpansion
\textbf{------------------------------------------------------}%
\begin{equation*}
f\ast g=\sum_{A,B,j,l,m}^{\infty }\left( i\hbar /2\right)
^{j}G_{j,l,m}^{A_{1}\cdots A_{l}B_{1}\cdots B_{m}}\left( D_{A_{1}}\cdots
D_{A_{l}}f\right) \left( D_{B_{1}}\cdots D_{B_{m}}g\right)
\end{equation*}%
%TCIMACRO{\TeXButton{\vspace{-15pt}}{\vspace{-15pt}}}%
%BeginExpansion
\vspace{-15pt}%
%EndExpansion

\qquad \qquad \qquad \qquad \qquad \qquad \qquad \qquad \qquad \qquad \qquad 
\textbf{------------------------------------------------------}

%TCIMACRO{\TeXButton{\begin{multicols}{2}}{\begin{multicols}{2}}}%
%BeginExpansion
\begin{multicols}{2}%
%EndExpansion
%TCIMACRO{\TeXButton{\vspace{-15pt}}{\vspace{-15pt}}}%
%BeginExpansion
\vspace{-15pt}%
%EndExpansion

for any two phase-space functions $f$ and $g$, where $D$ is a phase-space
connection and $G_{j,l,m}^{A_{1}\cdots A_{l}B_{1}\cdots B_{m}}$\ is an $l+m$
index tensor for each $j$, $l$, and $m$. See $\left[ \text{HiHe}\right] $
for conditions on these coefficients due to Gerstenhaber.

A star-product is a very complicated object because of the infinite order of
derivatives. In fact, it is this "infinite orderness" which is the source of
most of the trouble in working with star-products and why we need such fancy
tools in their construction and classification. Despite the difficult nature
of these objects, in the following thirty years, huge advances have been
achieved. One is the classification of star-products into equivalence
classes due to the contributions of many people (see $\left[ \text{DiSt}%
\right] $ for a brief history):

\textbf{Thm. }\textit{All star-products on a symplectic manifold (a
generalized phase-space) fall into equivalence classes which are
parametrized by a formal series in }$\hbar $,\textit{\ }$B=B_{0}+\hbar
B_{1}+\hbar ^{2}B_{2}+\cdots $\textit{\ with coefficients }$B_{i}$\textit{\
in the second de Rham cohomology group }$H_{dR}^{2}\left[ \left[ \hbar %
\right] \right] $\textit{.}

In each equivalence class, whether we describe our system with $\ast _{1}$
or $\ast _{2}$, all physical quantities (like expectation values) will be
identical. Two equivalent products $\ast _{1}$ and $\ast _{2}$ are related
by some invertable operator $T$:%
\begin{equation*}
T\left( f\right) =\sum_{A,j,l}^{\infty }\left( i\hbar /2\right)
^{j}T_{j,l}^{A_{1}\cdots A_{l}}\left( D_{A_{1}}\cdots D_{A_{l}}f\right)
\end{equation*}%
by the formula:%
\begin{equation*}
f\ast _{2}g=T^{-1}\left( T\left( f\right) \ast _{1}T\left( g\right) \right)
\end{equation*}

In simplified terms, this parametrization is a two-form $B$ that is closed ($%
dB=0$), but not necessarily exact ($B=dA$). It was shown in $\left[ \text{%
Bord1}\right] $ interpret $B$ as a magnetic field in our space-time. The
integral over any closed two-surface is the amount of magnetic monopole
charge sitting inside which is directly correlated to the fact that $B$ is
closed but not exact. For example, if the magnetic monopole charge in our
space-time is zero then all star-products are equivalent.

Another huge advance was developed by Fedosov in the construction of his
Fedosov star-product on an arbitrary finite-dimensional symplectic manifold
using geometric approaches. Fedosov, using insights from the index theorems
of Atiyah and others, perturbatively constructed a star-product via an
iterative process.$\left[ \text{GdT,Fed}\right] $ The convergence behavior
of the Fedosov star is still unknown in the general case, but some specific
exact formulas have been found.$\left[ \text{TiSp1,TiSp2}\right] $ With the
convergence issues aside, the classification theorem above means that 
\textit{all} star-products are equivalent to a Fedosov star. The properties
of the Fedosov star are $\left[ \text{Fed,TiSp2}\right] $:

\begin{enumerate}
\item 
%TCIMACRO{\TeXButton{\vspace{-15pt}}{\vspace{-15pt}}}%
%BeginExpansion
\vspace{-15pt}%
%EndExpansion
It is coordinate invariant.

\item It can be constructed on all symplectic manifolds (including all
phase-spaces) perturbatively in powers of $\hbar $.

\item It assumes no dynamics (e.g. Hamiltonian or Lagrangian), symmetries,
or even a metric.

\item The limit $\hbar \rightarrow 0$ yields classical mechanics.

\item It is equivalent to an operator formalism by a Weyl-like quantization
map.$\left[ \text{Fed,TiSp2}\right] $
\end{enumerate}

Another key development was the application of DQ to quantum field theory
(QFT) by Dito. In $\left[ \text{Dito}\right] $, Dito has successfully
constructed nonperturbative star-products for both a free covariant field
and a covariant field with a class of interaction terms which include
polynomial ones. The interacting case is done by using a cohomological
method of renormalization, called cohomological renormalization in which he
uses a linearization program. This involves identifying the diverging terms
as singular cocycles or coboundaries of the Hochschild cohomology in the
star-product which can be removed by changing to an equivalent star-product.

The main problem in DQ, as I see it, is related to the standard treatments
of deformation products which rely heavily on series expansions in a formal
parameter $\hbar $. Therefore, convergence of these series need to be
addressed which is a purpose of the results given here. Additionally, the
existence of a large enough set of states to describe physical systems
(which includes a notion of vaccum) needs to be addressed especially for
QFT's. Moreover, for a star-product on a Maxwell-Dirac field to be
constructed in a similar manner to that of $\left[ \text{Dito}\right] $, the
star-product of the asymptotic fields must be constructed (see $\left[ \text{%
HHS,HiHe2}\right] $ for work on fermions and bosons). Therefore, much work
still needs to be done in this area.

Additional important developments of DQ include the application to
statistical quantum mechanics where the KMS condition (a condition for a
state to be in thermodynamic equilibrium at a defined temperature, see $%
\left[ \text{BrRo,Haag}\right] $) was given in $\left[ \text{Bas}\right] $.
A formal definition of a KMS state of finitely many degrees of freedom was
defined in $\left[ \text{Bord2}\right] $. Also, a formal GNS construction of
a Hilbert space associated to any star-product has been formulated in $\left[
\text{BoWa}\right] $ and yields the correct results in the standard
representations such as the Bargmann and Schr\"{o}dinger representations.
Finally, Kontsevich in $\left[ \text{Kon}\right] $ has formulated a
star-product, called the Kontsevich star, on an arbitrary finite dimensional
Poisson manifold perturbatively in powers of $\hbar $.

*Note: The difficulty of constructing star-products is exemplified by Maxim
Kontsevich's Fields Medal in 1998, won in part because of his brilliant
construction of a star-product on arbitrary finite-dimensional Poisson
manifold called the Kontsevich star.$\left[ \text{Kon}\right] $ Furthermore,
this was the very first solution to a long-standing problem in mathematics:
showing that any finite dimensional Poisson manifold admits a formal
quantization.

%TCIMACRO{\TeXButton{\vspace{-15pt}}{\vspace{-15pt}}}%
%BeginExpansion
\vspace{-15pt}%
%EndExpansion

\section{Quantum Mechanics on Phase-Space: A 
%TCIMACRO{\TeXButton{\\}{\\ }}%
%BeginExpansion
\\ %
%EndExpansion
New Perspective of 
%TCIMACRO{\TeXButton{\\}{\\ }}%
%BeginExpansion
\\ %
%EndExpansion
Quantum Theory}

%TCIMACRO{\TeXButton{\vspace{-15pt}}{\vspace{-15pt}}}%
%BeginExpansion
\vspace{-15pt}%
%EndExpansion
The important question is: What advantage does deformation quantization (DQ)
have over the other standard formulations of quantum physics? Part of the
answer to this lies in the radically different framework: DQ is a theory of
quantum mechanics where observables are still phase-space functions.
Therefore, the conceptual break with classical mechanics is less severe than
the other two standard approaches.

Herein lies the advantage of DQ: Most techniques (geometrical or algebraic)
one uses with ordinary functions are valid in, or can be adapted in a
natural way into this framework. Therefore, DQ greatly expands our toolbox
including tools used in basic algebra and geometry. For instance, coordinate
invariance of quantum theory is easily manifest because the star product is
coordinate invariant. Also, different operator orderings in quantization can
be organized nicely into equivalence classes of star-products in DQ
parametrized by the formal series $B$. We note that there are examples of
equivalent orderings in $\left[ \text{Til,HiHe}\right] $.

Finally, building a theory restricted to a\ compact region in our
configuration space-time gives an innately local quantum theory. This can be
done by a simple restriction of set of all observables (which are just
functions in a formal series of $\hbar $) to a region of compact support. In
this sense one can build a local theory of quantum physics in an analogous
to AQFT.$\left[ \text{Haag,D\"{u}Fr}\right] $ Simply stated, the topology of
observable algebra is directly and naturally induced by the topology of the
algebra of functions on any given phase-space. Moreover, visualizing these
things require much less mental work because the physical observables are
the same functions they were in the classical theory.

This is contrary to the Hilbert space operator formulation in which
phase-space functions get mapped to fundamentally different objects. So much
on this set of operators is awkward and this awkwardness manifests itself in
the difficulty experienced when attempting some of the most basic things
that are easy in the classical theory. The reader may observe the extremely
complicated techniques to extend notions of topology, coordinate
transformations, derivatives, integrals, etc. into an operator setting in
noncommutative geometry using $\left[ \text{Con}\right] $ and in AQFT using $%
\left[ \text{Haag,BrRo}\right] $. The bulk of the hard work involved in DQ
is the construction of star-products, but once you have star-product many
other things are a lot easier. Therefore, if these star-products exist for a
particular system, it seems useful to formulate or reformulate them using DQ.

%TCIMACRO{\TeXButton{\vspace{-15pt}}{\vspace{-15pt}}}%
%BeginExpansion
\vspace{-15pt}%
%EndExpansion

\section{The Klein-Gordon 
%TCIMACRO{\TeXButton{\\}{\\ }}%
%BeginExpansion
\\ %
%EndExpansion
Equation}

%TCIMACRO{\TeXButton{\vspace{-15pt}}{\vspace{-15pt}}}%
%BeginExpansion
\vspace{-15pt}%
%EndExpansion
It has been shown in $\left[ \text{TiSp1,TiSp2,Til}\right] $ that the
Klein-Gordon (KG) equation in an arbitrary (possibly curved) space-time may
formulated using Fedosov's Weyl-like quantization map $\sigma ^{-1}$ (i.e.
the analogue of $\mathcal{W}$) and its inverse $\sigma $ as:%
\begin{equation}
H\ast \rho _{m}=\rho _{m}\ast H=m^{2}\rho _{m}  \label{DQKG}
\end{equation}%
\begin{equation}
H=p_{\mu }\ast p^{\mu }+\xi R\left( x\right)  \label{H}
\end{equation}%
\begin{equation*}
Tr_{\ast }\left( \rho _{m}\right) =1~~~,~~~\bar{\rho}_{m}=\rho _{m}
\end{equation*}%
where $\ast $ is the Fedosov\ star-product, $R\left( x\right) $ is the Ricci
scalar, $p^{\mu }:=g^{\mu \nu }p_{\nu }$, and $\xi \in 
%TCIMACRO{\U{2102} }%
%BeginExpansion
\mathbb{C}
%EndExpansion
$ is an arbitrary constant.$\left[ \text{Fed,GdT}\right] $ Also, $\rho _{m}$
is called a Wigner function and is defined by:

\begin{equation*}
\rho _{m}:=\sigma \left( \left\vert \phi _{m}\rangle \langle \phi
_{m}\right\vert \right)
\end{equation*}%
In the case of the dS and AdS space-times given by the embedding:%
\begin{equation*}
\eta _{\mu \nu }x^{\mu }x^{\nu }=1/C\text{ \ \ and \ \ }x^{\mu }p_{\mu }=A
\end{equation*}%
Omitting the technical details, we obtain the exact commutators for the
Fedosov star-product:%
\begin{equation}
\left[ x^{\mu },x^{\nu }\right] _{\ast }=0~~\ ~~[x_{\mu },M_{\nu \rho
}]_{\ast }=i\hbar x_{[\nu }\eta _{\rho ]\mu }
\end{equation}%
\begin{equation*}
\lbrack M_{\mu \nu },M_{\rho \sigma }]_{\ast }=i\hbar (M_{\rho \lbrack \mu
}\eta _{\nu ]\sigma }-M_{\sigma \lbrack \mu }\eta _{\nu ]\rho })
\end{equation*}%
indices run from $0$ to $4$, $M_{\mu \nu }=x_{[\mu }\ast p_{\nu ]}$, $x_{\mu
}=\eta _{\mu \nu }x^{\nu }$.

The conditions of the embedding $x^{\mu }x_{\mu },~x^{\mu }p_{\mu }=x^{\mu
}\ast p_{\mu }$ become the Casimir invariants of the algebra in group
theoretic language.

We now summarize our two key observations:

\begin{enumerate}
\item 
%TCIMACRO{\TeXButton{\vspace{-10pt}}{\vspace{-10pt}}}%
%BeginExpansion
\vspace{-10pt}%
%EndExpansion
$M$'s generate $\mathbb{SO}\left( 1,4\right) $ and $\mathbb{SO}\left(
2,3\right) $ for dS and AdS respectively.

\item $M$'s and $x$'s generate $\mathbb{SO}\left( 2,4\right) $ for \textit{%
both} dS and AdS.
\end{enumerate}

%TCIMACRO{\TeXButton{\vspace{-10pt}}{\vspace{-10pt}}}%
%BeginExpansion
\vspace{-10pt}%
%EndExpansion
By calculating $R=-4C$ and $p_{\mu }\ast p^{\mu }$ in terms of $M$ and $x$,
the Hamiltonian $\left( \text{%
%TCIMACRO{\TeXButton{\ref{H}}{\ref{H}}}%
%BeginExpansion
\ref{H}%
%EndExpansion
}\right) $ is:%
\begin{equation}
H=2CM_{\mu \nu }\ast M^{\mu \nu }+\left( A-4i\hbar \right) AC-4\xi C
\end{equation}%
where $M_{\mu \nu }\ast M^{\mu \nu }$ is a Casimir invariant of the subgroup 
$\mathbb{SO}\left( 1,4\right) $ or $\mathbb{SO}\left( 2,3\right) $ for dS or
AdS respectively.

In the more familiar form of Hilbert space language the KG equation takes
the form:%
\begin{equation}
(2C\hat{M}_{\mu \nu }\hat{M}^{\mu \nu }+\chi C)\left\vert \phi
_{m}\right\rangle =m^{2}\left\vert \phi _{m}\right\rangle
\end{equation}%
where $\left\langle \phi _{m}|\phi _{m}\right\rangle =1$, $%
%TCIMACRO{\U{2102} }%
%BeginExpansion
\mathbb{C}
%EndExpansion
\ni \chi =\left( A-4i\hbar \right) A-4\xi $ is an arbitrary constant, and we
regard all groups to be in a standard irreducible representation on the set
of linear Hilbert space operators.

*Note: This result is confirmed by $\left[ \text{Fro1-4}\right] $ as well as
others.

%TCIMACRO{\TeXButton{\vspace{-15pt}}{\vspace{-15pt}}}%
%BeginExpansion
\vspace{-15pt}%
%EndExpansion

\section{Covariant Free Field Quantization and the Dito Star-Product}

%TCIMACRO{\TeXButton{\vspace{-15pt}}{\vspace{-15pt}}}%
%BeginExpansion
\vspace{-15pt}%
%EndExpansion
According to Feynman, in QFT positive frequency (energy) solutions to the KG
equation correspond to particles that moving forward in time while negative
ones correpond to particles moving backwards in time. Particles moving
backwards in time correspond to anti-particles moving forward in time. This
is why Fourier modes are most suitable for QFT (see $\left[ \text{HiHe}%
\right] $ for more details).

Given the free KG equation:%
\begin{equation*}
\left( \partial _{\mu }\partial ^{\mu }-m^{2}/\hbar ^{2}\right) \phi \left( 
\mathbf{x},t\right) =0
\end{equation*}%
on Minkowski space. First we decompose initial data $\Phi \left( \mathbf{x}%
,0\right) :=\left( \phi \left( \mathbf{x},0\right) ,\pi \left( \mathbf{x}%
,0\right) \right) $ into Fourier modes of definite energy:%
\begin{equation*}
\phi \left( \mathbf{x},0\right) =\int_{\Sigma }\frac{d^{3}\mathbf{k}}{%
2\left( 2\pi \right) ^{3/2}\omega \left( \mathbf{k}\right) }\left( \bar{a}_{%
\mathbf{k}}e^{-i\mathbf{k}\cdot \mathbf{x}}+a_{\mathbf{k}}e^{i\mathbf{k}%
\cdot \mathbf{x}}\right)
\end{equation*}%
\begin{equation*}
\pi \left( \mathbf{x},0\right) =i\int_{\Sigma }\frac{d^{3}\mathbf{k}}{%
2\left( 2\pi \right) ^{3/2}}\left( \bar{a}_{\mathbf{k}}e^{-i\mathbf{k}\cdot 
\mathbf{x}}-a_{\mathbf{k}}e^{i\mathbf{k}\cdot \mathbf{x}}\right)
\end{equation*}%
where $\pi \left( \mathbf{x}\right) =\partial _{t}\phi \left( \mathbf{x}%
\right) $ and $\omega \left( \mathbf{k}\right) :=\left( \mathbf{k}%
^{2}+m^{2}\right) ^{1/2}$.

The set of solutions $\Phi \left( \mathbf{x},t\right) :=\left( \phi \left( 
\mathbf{x},t\right) ,\pi \left( \mathbf{x},t\right) \right) $ to the KG
equation is a (infinite-dimensional) Poisson manifold with Poisson structure:

$\left[ \Phi ,\Psi \right] _{P}$%
\begin{equation*}
=\frac{2}{i}\int_{\Sigma }d^{3}\mathbf{k}\left( D_{a_{\mathbf{k}}}\left(
\Phi \right) D_{\bar{a}_{\mathbf{k}}}\left( \Psi \right) -D_{\bar{a}_{%
\mathbf{k}}}\left( \Phi \right) D_{a_{\mathbf{k}}}\left( \Psi \right) \right)
\end{equation*}%
\begin{equation*}
D_{a_{\mathbf{k}}}=\sqrt{\left( 2\omega \left( \mathbf{k}\right) \right) }%
\frac{\delta }{\delta a_{\mathbf{k}}}~~~~,~~~\ D_{\bar{a}_{\mathbf{k}}}=%
\sqrt{\left( 2\omega \left( \mathbf{k}\right) \right) }\frac{\delta }{\delta 
\bar{a}_{\mathbf{k}}}
\end{equation*}%
where $\Phi =\Phi \left( \mathbf{x},t\right) $ and $\Psi \left( \mathbf{x}%
,t\right) $ are any two solutions of the KG equation and the Poisson bracket
is invariant under choice of hypersurface $\Sigma $.$\left[ \text{HaEl}%
\right] $

Just as in the case of ordinary quantum mechanics, this Poisson structure
induces a star-product on the phase-space. Because different star-products
correspond to different operator orderings we must be careful with the
star-product we choose to prevent large numbers of divergences.

The choice in Dito's star product for a free field is normal ordering
because normal ordering plays a special role in QFT by annihilating the
vacuum state. This choice eliminates the artificial infinity in the vacuum
energy of the free field. The normal star-product is defined by:%
\begin{equation*}
\Phi \ast _{N}\Psi =\Phi \Psi +\sum_{n=1}^{\infty }\hbar ^{n}C_{n}^{N}\left(
\Phi ,\Psi \right)
\end{equation*}%
where:

%TCIMACRO{\TeXButton{\end{multicols}}{\end{multicols}}}%
%BeginExpansion
\end{multicols}%
%EndExpansion
%TCIMACRO{\TeXButton{\vspace{-15pt}}{\vspace{-15pt}}}%
%BeginExpansion
\vspace{-15pt}%
%EndExpansion
\textbf{------------------------------------------------------}%
\begin{equation*}
C_{n}^{N}\left( \Phi ,\Psi \right) :=\frac{1}{n!}\int d^{3}\mathbf{k}%
_{1}\cdots d^{3}\mathbf{k}_{n}\left( D_{a_{\mathbf{k}_{1}}}\cdots D_{a_{%
\mathbf{k}_{n}}}\left( \Phi \right) D_{\bar{a}_{\mathbf{k}_{1}}}\cdots D_{%
\bar{a}_{\mathbf{k}_{n}}}\left( \Psi \right) \right)
\end{equation*}%
%TCIMACRO{\TeXButton{\vspace{-15pt}}{\vspace{-15pt}}}%
%BeginExpansion
\vspace{-15pt}%
%EndExpansion

\qquad \qquad \qquad \qquad \qquad \qquad \qquad \qquad \qquad \qquad \qquad 
\textbf{------------------------------------------------------}

%TCIMACRO{\TeXButton{\begin{multicols}{2}}{\begin{multicols}{2}}}%
%BeginExpansion
\begin{multicols}{2}%
%EndExpansion
%TCIMACRO{\TeXButton{\vspace{-15pt}}{\vspace{-15pt}}}%
%BeginExpansion
\vspace{-15pt}%
%EndExpansion

For an interacting scalar field on Minkowski space with, for example, a
polynomial interaction term $V\left( \phi \right) $ the KG equation is $%
\left( \partial _{\mu }\partial ^{\mu }-m^{2}/\hbar ^{2}+V^{\prime }\left(
\phi \right) \right) \phi \left( \mathbf{x},t\right) =0$. Through a
linearization program, Dito constructed star-product $\ast ^{+}$ at the
asymptotic future free field and $\ast ^{-}$ at the asymptotic past free
field which is free of a large number of divergences by construction.

By investigating infinite dimensional star-products associated to
interacting fields Dito saw how divergences form and how renormalization
should be performed to remove them. The diverging terms are singular
cocycles or coboundaries of the Hochschild cohomology in the star-product
which can be removed by changing to an equivalent star-product which is
achieved by the linearization program.

For a star-product on a Maxwell-Dirac field to be constructed in a similar
manner, the star-product of the asymptotic fields must be constructed.
Recently in $\left[ \text{HiHe2}\right] $ a star-product was constructed for
a fermionic system. This is based on the works of Berezin and Marinov in
which they begin with a pseudoclassical system based on Grassmann algebras.
Hirshfeld and Hensler proposed a fermionic star-product as a deformation
quantization of this pseudoclassical system. However, work still needs to be
done to understand how to properly construct the asymptotic fields of QED in
DQ.

%TCIMACRO{\TeXButton{\vspace{-15pt}}{\vspace{-15pt}}}%
%BeginExpansion
\vspace{-15pt}%
%EndExpansion

\section{Connections to the 
%TCIMACRO{\TeXButton{\\}{\\ }}%
%BeginExpansion
\\ %
%EndExpansion
Algebraic Approach to Quantum Field Theory}

%TCIMACRO{\TeXButton{\vspace{-15pt}}{\vspace{-15pt}}}%
%BeginExpansion
\vspace{-15pt}%
%EndExpansion
The Algebraic Approach to quantum field theory (AQFT), invented by Rudolf
Haag and Daniel Kastler, formulates QFT from a sufficiently rigorous
axiomatic framework which is consistent with all the basic principles of QFT.%
$\left[ \text{Buc,Haag,BrRo}\right] $ This is contrast to the alternative
approach of Constructive QFT which builds on existing mathematical methods
for the treatment of physical models.

The relationship between conventional approaches to QFT's such as
Lagrangian/path integral formulations and canonical quantization to AQFT may
be compared to the relationship between the coordinate dependent approach to
differential geometry and the coordinate independent one.$\left[ \text{Buc}%
\right] $ To calculate certain quantities it is natural (and sometimes
necessary) to describe the situation with coordinates, e.g. writing down
Christoffel symbols, etc. However, if you are more interested in a general
abstract analysis of the manifold you need coordinate independent quantities
like a connection, fiber and vector bundles, etc. It is only through the use
of the two complementary approaches can a full understanding of differential
geometry be yielded. In this way, a rigorous axiomatic framework for QFT is
necessary to compliment the other two methods. AQFT attempts to achieve this
more rigorous framework.

The advances obtained through AQFT include the clarification of the roles of
locality and covariance in QFT. Also, in AQFT, equivalent QFT's have the
same abstract AQFT structure. However, the issue of the existence of a
suitable set of states (which includes a notion of a vacuum) is a difficult
and ongoing debate except for the class of maximally symmetric spaces which
has been solved. Additionally, there have been numerous successes on an
arbitrary globally hyperbolic space-time in perturbative AQFT using
renormalization techniques such as configuration space and micro-local
techniques. The main new insight here is the complete disentanglement of the
ultraviolet and infrared problems in the perturbative expansion. Also,
insights into holography of the AdS/CFT correspondence (triggered by string
theory) have been made in AQFT. Holography is the notion that a QFT in the
bulk of a manifold uniquely determines the QFT on the boundary.

As is pointed out in $\left[ \text{D\"{u}Fr}\right] $, on the level of
concrete models AQFT was less successful. It is in this regard that we
believe that DQ can come to the rescue by defining a \textit{concrete} and 
\textit{practical}\ phase-space description of an abstract net of
observables (the standard set used in AQFT). It was conjectured in $\left[ 
\text{D\"{u}Fr}\right] $ that the relationship between perturbative AQFT may
be understood in terms of DQ (also see $\left[ \text{HiHe1}\right] $). A
reason is because the DQ observables are ordinary functions; therefore, a
local theory can be built out of functions of compact support in a way that
is entirely natural. This set of observables is the "net of observables"
described in AQFT. As long as the set of states that are well-defined in DQ
is large enough to describe the theory as well as the convergence of all
relevant quantities, DQ should be very fruitful. In addition, the
conventional treatment of DQ in terms of a formal series in $\hbar $ is
extremely convenient for perturbative calculations.

%TCIMACRO{\TeXButton{\vspace{-15pt}}{\vspace{-15pt}}}%
%BeginExpansion
\vspace{-15pt}%
%EndExpansion

\section{Conclusion}

%TCIMACRO{\TeXButton{\vspace{-15pt}}{\vspace{-15pt}}}%
%BeginExpansion
\vspace{-15pt}%
%EndExpansion
What we have in DQ is a coordinate and dynamics independent formulation of
quantum theory formulated on phase-space.$\left[ \text{Til,GoRe}\right] $
DQ's conceptual advantages come from the fact that the observables are
phase-space functions just as they are in classical mechanics$.$ Therefore,
many geometric and algebraic tools can be extended very naturally into DQ.
It seems that most of the hard work in DQ is in the construction of
star-products and issues relating to their convergence. However, once you
have a star-product many things like topology, coordinate transformations,
etc. are much easier to understand on a fundamentally conceptual level.

Advances have been made in applying DQ to QFT's. The star-products of Dito
on scalar fields, including those with polynomial interaction terms are some
of the significant advances.$\left[ \text{Dito}\right] $ In addition, their
seems to be a strong connection between DQ and AQFT because of the
similarities between the two.$\left[ \text{D\"{u}Fr}\right] $ One includes
their focus on the algebra of observables. Another is the "net of
observables" in AQFT can be constructed in DQ by the simple restriction of
the observables which are functions on the phase-space. The reason we should
care about this connection is because AQFT\ has provided many advances in
the understanding of QFT's. These include the role of locality, the
disentanglement of the ultraviolet and infrared problems on an arbitrary
globally hyperbolic space-time, and insights to the AdS/CFT correspondance.
We hope that DQ may be able to provide the concrete models that AQFT lacks.

By applying DQ to QFT's we may hopefully yield a better understanding of
quantum theory.

%TCIMACRO{\TeXButton{\vspace{-15pt}}{\vspace{-15pt}}}%
%BeginExpansion
\vspace{-15pt}%
%EndExpansion

\section*{Acknowledgements}

%TCIMACRO{\TeXButton{\vspace{-15pt}}{\vspace{-15pt}}}%
%BeginExpansion
\vspace{-15pt}%
%EndExpansion
I would like to thank my advisor George Sparling for very helpful
discussions.

\section*{References}

%TCIMACRO{\TeXButton{\vspace{-10pt}}{\vspace{-10pt}}}%
%BeginExpansion
\vspace{-10pt}%
%EndExpansion
$\left[ \text{Bord1}\right] $ Bordemann et al., \textit{On representations
of star product algebras over cotangent spaces on Hermitian line bundle},
math.QA/9811055v2.

%TCIMACRO{\TeXButton{\vspace{-10pt}}{\vspace{-10pt}}}%
%BeginExpansion
\vspace{-10pt}%
%EndExpansion
$\left[ \text{Buc}\right] $ Buchholz, D. Algebraic Quantum Field Theory: A
Status Report, talk given at the XIIIth International Congress on
Mathematical Physics, London (2000), math-ph/0011044.

%TCIMACRO{\TeXButton{\vspace{-10pt}}{\vspace{-10pt}}}%
%BeginExpansion
\vspace{-10pt}%
%EndExpansion
$\left[ \text{BrRo}\right] $ Brattelli O., Robinson D. W.; Operator Algebras
and Quantum Statistical Mechanics I. and II, New York, Springer-Verlang
(1979).

%TCIMACRO{\TeXButton{\vspace{-10pt}}{\vspace{-10pt}}}%
%BeginExpansion
\vspace{-10pt}%
%EndExpansion
$\left[ \text{Con}\right] $ Connes A., Noncommutative Geometry, San Diego:
Academic Press (1994).

%TCIMACRO{\TeXButton{\vspace{-10pt}}{\vspace{-10pt}}}%
%BeginExpansion
\vspace{-10pt}%
%EndExpansion
$\left[ \text{HaEl}\right] $ Hawking S., Ellis G.; The Large Scale Structure
of Space-Time, Cambridge University Press (1973).

%TCIMACRO{\TeXButton{\vspace{-10pt}}{\vspace{-10pt}}}%
%BeginExpansion
\vspace{-10pt}%
%EndExpansion
$\left[ \text{Dito}\right] $ Dito, G.; Deformation Quantization of Covariant
Fields, IRMA Lectures in Math. Theoret. Phys. 1; Deformation Quantization,
Proceedings of the Meeting of Theoretical Physicists and Mathematicians,
Strasbourg 2001 (G.Halbout ed.); Walter de Gruyter, Berlin 2001,
math.QA/0202271v1.

%TCIMACRO{\TeXButton{\vspace{-10pt}}{\vspace{-10pt}}}%
%BeginExpansion
\vspace{-10pt}%
%EndExpansion
$\left[ \text{DiSt}\right] $ Dito G., Sternheimer D.; Deformation
Quantization; Genesis, Developments and Metamorphoses, math.QA/0201168v1.

%TCIMACRO{\TeXButton{\vspace{-10pt}}{\vspace{-10pt}}}%
%BeginExpansion
\vspace{-10pt}%
%EndExpansion
$\left[ \text{D\"{u}Fr}\right] $ D\"{u}tsch M., Fredenhagen K.; Perturbative
Algebraic Field Theory, and Deformation Quantization, hep-th/0101079v1.

%TCIMACRO{\TeXButton{\vspace{-10pt}}{\vspace{-10pt}}}%
%BeginExpansion
\vspace{-10pt}%
%EndExpansion
$\left[ \text{Fed}\right] $ Fedosov\ B.;\textit{\ }Deformation Quantization
and Index Theory, Akademie, Berlin 1996.

%TCIMACRO{\TeXButton{\vspace{-10pt}}{\vspace{-10pt}}}%
%BeginExpansion
\vspace{-10pt}%
%EndExpansion
$\left[ \text{Fro1}\right] $ C. Fr\o nsdal, \textit{Elementary Particles in
a Curved Space}, Rev. Mod. Phys. \textbf{37}, I, 221 (1965).

%TCIMACRO{\TeXButton{\vspace{-10pt}}{\vspace{-10pt}}}%
%BeginExpansion
\vspace{-10pt}%
%EndExpansion
$\left[ \text{Fro2}\right] $ C. Fr\o nsdal, \textit{Elementary Particles in
a Curved Space II}, Phys. Rev. D Vol. \textbf{10} Num. 2, 589 (1973).

%TCIMACRO{\TeXButton{\vspace{-10pt}}{\vspace{-10pt}}}%
%BeginExpansion
\vspace{-10pt}%
%EndExpansion
$\left[ \text{Fro3}\right] $ C. Fr\o nsdal, \textit{Elementary Particles in
a Curved Space III}, Phys. Rev. D Vol. \textbf{12} Num. 12, 3810 (1975).

%TCIMACRO{\TeXButton{\vspace{-10pt}}{\vspace{-10pt}}}%
%BeginExpansion
\vspace{-10pt}%
%EndExpansion
$\left[ \text{Fro4}\right] $ C. Fr\o nsdal, \textit{Elementary Particles in
a Curved Space IV}, Phys. Rev. D Vol. \textbf{12} Num. 12, 3819 (1975).

%TCIMACRO{\TeXButton{\vspace{-10pt}}{\vspace{-10pt}}}%
%BeginExpansion
\vspace{-10pt}%
%EndExpansion
$\left[ \text{Giu}\right] $ Giulini D.; That Strange Procedure Called
Quantizaiton, quant-ph/0304202.

%TCIMACRO{\TeXButton{\vspace{-10pt}}{\vspace{-10pt}}}%
%BeginExpansion
\vspace{-10pt}%
%EndExpansion
$\left[ \text{GoRe}\right] $ Gozzi E., Reuter M.; Quantum Deformed Canonical
Transformations $W_{\infty }$-Algebras and Unitary Transformations,
hep-th/0306221v1.

%TCIMACRO{\TeXButton{\vspace{-10pt}}{\vspace{-10pt}}}%
%BeginExpansion
\vspace{-10pt}%
%EndExpansion
$\left[ \text{Gro}\right] $ Groenewold H. J.; On the Principles of
Elementary Quantum Mechanics, Physica \textbf{12} (1946), 405--460.

%TCIMACRO{\TeXButton{\vspace{-10pt}}{\vspace{-10pt}}}%
%BeginExpansion
\vspace{-10pt}%
%EndExpansion
$\left[ \text{GdT}\right] $\ Gadella M., del Olmo M.A., Tosiek J.;\textit{\ }%
Geometrical Origin of the $\ast -$product in the Fedosov Formalism\textit{,}
hep-th/0405157v1.

%TCIMACRO{\TeXButton{\vspace{-10pt}}{\vspace{-10pt}}}%
%BeginExpansion
\vspace{-10pt}%
%EndExpansion
$\left[ \text{Haag}\right] $ Haag R.; Local Quantum Physics,
Springer-Verlang (1992).

%TCIMACRO{\TeXButton{\vspace{-10pt}}{\vspace{-10pt}}}%
%BeginExpansion
\vspace{-10pt}%
%EndExpansion
$\left[ \text{HWW}\right] $ Hancock\ J., Walton M., Wynder B.; Quantum
Mechanics Another Way, physics/0405029v1.

%TCIMACRO{\TeXButton{\vspace{-10pt}}{\vspace{-10pt}}}%
%BeginExpansion
\vspace{-10pt}%
%EndExpansion
$\left[ \text{HiHe}\right] $ Hirshfeld A., Henselder P.; Deformation
Quantization in the Teaching of Quantum Mechanics, Am. J. Phys. \textbf{70}
(5), May 2002, quant-ph/0208163.

%TCIMACRO{\TeXButton{\vspace{-10pt}}{\vspace{-10pt}}}%
%BeginExpansion
\vspace{-10pt}%
%EndExpansion
$\left[ \text{Kon}\right] $ Kontsevich M.; Letters in Mathematical Physics 
\textbf{66}, 157 (2003), q-alg/9709040v1.

%TCIMACRO{\TeXButton{\vspace{-10pt}}{\vspace{-10pt}}}%
%BeginExpansion
\vspace{-10pt}%
%EndExpansion
$\left[ \text{Moy}\right] $ Moyal J.E.; Quantum mechanics as a statistical
theory, Proceedings of the Cambridge Philosophical Society, \textbf{45}
(1949), 99-124.

%TCIMACRO{\TeXButton{\vspace{-10pt}}{\vspace{-10pt}}}%
%BeginExpansion
\vspace{-10pt}%
%EndExpansion
$\left[ \text{TiSp1}\right] $ Tillman P., Sparling G.; Fedosov Observables
on Constant Curvature Manifolds and the Klein-Gordon Equation, gr-qc/0603017.

%TCIMACRO{\TeXButton{\vspace{-10pt}}{\vspace{-10pt}}}%
%BeginExpansion
\vspace{-10pt}%
%EndExpansion
$\left[ \text{TiSp2}\right] $ Tillman P., Sparling G.; Observables of
Angular Momentum as Observables on the Fedosov Quantized Sphere, J. Math.
Phys. \textbf{47}, 052102 (2006), quant-ph/0509210.

%TCIMACRO{\TeXButton{\vspace{-10pt}}{\vspace{-10pt}}}%
%BeginExpansion
\vspace{-10pt}%
%EndExpansion
$\left[ \text{Til}\right] $ Tillman P., Deformation Quantization,
Quantization, and the Klein-Gordon Equation, \textit{Proceedings from The
Second International Conference on Quantum Theories and Renormalization
Group in Gravity and Cosmology}, \textit{Preprint} gr-qc/0610141.

%TCIMACRO{\TeXButton{\vspace{-10pt}}{\vspace{-10pt}}}%
%BeginExpansion
\vspace{-10pt}%
%EndExpansion
$\left[ \text{Wal1}\right] $ Waldmann S.; On the Representation Theory of
Deformation Quantization, math.QA/0107112v2.

%TCIMACRO{\TeXButton{\vspace{-10pt}}{\vspace{-10pt}}}%
%BeginExpansion
\vspace{-10pt}%
%EndExpansion
$\left[ \text{Wal2}\right] $ Waldmann S.; States and Representations in
Deformation Quantization, QA/0408217v1.

%TCIMACRO{\TeXButton{\vspace{-10pt}}{\vspace{-10pt}}}%
%BeginExpansion
\vspace{-10pt}%
%EndExpansion
$\left[ \text{Weyl}\right] $ Weyl H.; The Theory of Groups and Quantum
Mechanics, Dover, New-York 1931, translated from Gruppentheorie und
Quantenmechanik, Hirzel Verlag, Leipzig 1928; \textquotedblleft
Quantenmechanik und Gruppentheorie,\textquotedblright\ Z. Physik \textbf{46}
(1927), 1--46.

%TCIMACRO{\TeXButton{\vspace{-10pt}}{\vspace{-10pt}}}%
%BeginExpansion
\vspace{-10pt}%
%EndExpansion
$\left[ \text{Wig}\right] $ Wigner E.P.; Quantum Corrections for
Thermodynamic Equilibrium, Phys. Rev. \textbf{40} (1932), 749--759.

%TCIMACRO{\TeXButton{\vspace{-10pt}}{\vspace{-10pt}}}%
%BeginExpansion
\vspace{-10pt}%
%EndExpansion
$\left[ \text{ZaCu}\right] $ Zachos C., Curtright T.; Phase-space
Quantization of Field Theory, ANL-HEP-CP-99-06 Miami TH/1/99,
hep-th/9903254v2.

%TCIMACRO{\TeXButton{\end{multicols}}{\end{multicols}}}%
%BeginExpansion
\end{multicols}%
%EndExpansion

\end{document}